\documentclass[twocolumn,showpacs,preprintnumbers,amsmath,amssymb,superscriptaddress]{revtex4}

\usepackage{graphicx}
\usepackage{dcolumn}
\usepackage{bm}
\usepackage{amsmath}

\begin{document}

\preprint{LA-UR-06-2202}

\title{Neutrino signatures of supernova turbulence}

\author{Alexander Friedland}
 \email{friedland@lanl.gov}
\affiliation{%
Theoretical Division, T-8, MS B285, Los Alamos National Laboratory, Los Alamos, NM 87545
}%
\author{Andrei Gruzinov}
 \email{andrei@physics.nyu.edu}
\affiliation{%
New York University, 4 Washington Place, New York, NY 10003
}%

\date{July 11, 2006}

\begin{abstract}

Convection that develops behind the shock front during the first
second of a core-collapse supernova explosion is believed to play
a crucial role in the explosion mechanism. We demonstrate that the
resulting turbulent density fluctuations may be directly observable
in the neutrino signal starting at $t\gtrsim 3-4$ s after the onset of
the explosion. The effect comes from the modulation of the MSW
flavor transformations by the turbulent density fluctuations. We
derive a simple and general criterion for neutrino flavor
depolarization in a Kolmogorov-type turbulence and apply it to the
turbulence seen in modern numerical simulations. The turbulence
casts a ``shadow", by making other features, such as the shock
front, unobservable in the density range covered by the
turbulence.

\end{abstract}

\pacs{14.60.Pq, 97.60.Bw, 97.10.Cv, 42.25.Dd}
\maketitle

\section{Introduction}
The leading proposal for the mechanism of core-collapse supernova
explosions is that
the shock first stalls and then regains its energy through heating
of the material behind it by streaming neutrinos
\cite{ColgateWhite1966,Wilson1985}.
Within this paradigm a crucial role is played by vigorous
convection behind the shock front
\cite{Bethe1990,HerantColgate1994}. This convection, clearly seen
in modern multidimensional simulations
\cite{FryerWarren2003,JankaPRL,Kifonidis:2005yj,Scheck:2006rw,Burrowsacoustic},
brings the energy deposited by neutrinos in dense regions
to the region behind the stalled shock front (for a recent
overview and references see, e.g., \cite{WoosleyJanka}). Besides
the shock revival, the convection leads to asymmetric accretion
onto the protoneutron star, providing a compelling explanation for
the high observed pulsar velocities \cite{JankaPRL}. Finally,
evidence for the convection comes from observations of SN1987A
that indicate extensive mixing during the early stages of the
explosion \cite{HerantBenz1992}.

Turbulent convective motions create a fluctuating density field in
the post-shock region. Importantly, these fluctuations  remain
long after the shock restarts and a successful explosion is
obtained \cite{Kifonidis:2005yj,Scheck:2006rw}. In particular,
they persist over the duration of the neutrino burst ($\sim 10$
sec) and can thus modulate the MSW flavor transformations of the
neutrinos. The goal of the present work is to show that the
turbulence seen in the simulations indeed leaves an imprint on the
neutrino signal, possibly starting from about 3-4 seconds after
the onset of the explosion. This imprint may replace, or combine
with, signatures of other features already discussed in the
literature, such as the passage of the front \cite{SchiratoFuller}
and reverse \cite{RaffeltDighe} shocks through the resonance layer
(see also \cite{Lisi2004}), or effects of low-density bubbles
\cite{KnellerMcLaughlin}.

\section{Outline}
A core-collapse supernova emits both neutrinos and antineutrinos,
of all three flavors, with energies $\sim$10-20 MeV. The spectra
and luminosities for $\nu_e$, $\bar\nu_e$, and $\nu_x$
($x=\mu,\tau,\bar\mu,\bar\tau$) are in general different.
On the way out of the star, the neutrinos undergo matter-enhanced
flavor transformations at certain characteristic (``resonant")
densities, $\rho_H\sim 2.4\times 10^{3}$ g/cm$^3$ $(15\mbox{
MeV}/E_\nu)(0.5/Y_e)$ ($H$ or ``high" resonance) and $\rho_L\sim
\mbox 7\times10^{1}$ g/cm$^3$ $(15\mbox{ MeV}/E_\nu)(0.5/Y_e)$
($L$ or ``low" resonance) \footnote{These densities are set by the
mass splittings measured by the solar/reactor neutrino and
atmospheric/beam experiments correspondingly: $\Delta
m_\odot=8\times 10^{-5}$ eV$^2$
\cite{KamLANDspectrum2004} and $\Delta m_{atm}=2.7\times 10^{-3}$
eV$^2$ (the best fit to the atmospheric/K2K/MINOS data
\cite{myMINOS}). Notice that factors of $\cos2\theta$, which
traditionally appear in the definition of the resonance, lead to
physically incorrect conclusions for large mixing angles
\cite{myresonance}. In particular, for the $L$ resonance the
traditional definition implies that the flavor transformation
should happen either in the neutrino or antineutrino channel,
while in reality significant transformation occurs in both.}.
These transformations permute the neutrinos in various flavors and
determine the neutrino spectra detected on Earth.

To estimate the effect of the turbulence on the permutation
efficiency, we need to know the spectrum of the density
fluctuations in the turbulence and the response of neutrino
evolution to stochastic fluctuations of different sizes. The two
tasks are tightly coupled. The fluctuations relevant to neutrino
evolution turn out to be smaller than the resolution of the
present simulations and, hence,  must be inferred from the
large-scale features seen in the simulations by physical
arguments. At the same time, the spectrum of fluctuations in a
physical turbulence is very different from the
``$\delta$-correlated" noise for which analytical solutions for
neutrino evolution exist
\cite{Nicolaidis,Loreti1994,Loreti:1995ae,Balantekin1996,BurgessMichaud,BurgessProceedings}.
The appropriate treatment will be developed here.

\section{Turbulence}
As we will see, the scales of interest are $\lambda\lesssim
15/\sin2\theta_{13}$ km. Unless $\sin2\theta<10^{-3}$, these are
much smaller than the local scale height $r$ at $t\gtrsim4$ sec.
Due to Rayleigh-Taylor and Kelvin-Helmholts instabilities, which
are clearly seen in the simulations, Kolmogorov cascade into
smaller scales should develop. Density (just like velocity) will
then exhibit power-law fluctuations on all inertial range scales,
that is on all scales between $r$ and viscous cut-off $l_v$
($l_v\ll\lambda\ll r$),
%
\begin{eqnarray}
 \label{eq:Cpowerlaw}
C(k) \equiv\int dx \langle \delta n(0)\delta n(x)\rangle
e^{-ikx}=C_0 k^{\alpha}.
\end{eqnarray}
For the Kolmogorov turbulence the Fourier transform of the velocity
correlator has $\alpha=-5/3$. We will consider the range
$-2\lesssim\alpha\lesssim-1.5$.

Eq.~(\ref{eq:Cpowerlaw}) implies that a density variation between
two points separated by a distance $\lambda$ scales as
$\sim\left[\int_{2\pi/\lambda}^{2\pi/l_{v}} \frac{dk}{2\pi} C_0
k^{\alpha}\right]^{1/2}$. For $\alpha>-1$ the variation has a
strong unphysical dependence on the UV cutoff, $\propto
l_{v}^{-(\alpha+1)/2}$, and diverges as $l_{v}\rightarrow0$. This
is obvious for the $\delta$-correlated noise, $\langle\delta
n_e(x) \delta n_e(y) \rangle =const \times \delta(x-y)$, for which
$\alpha=0$: the small-scale variation is infinite.
Even if one tries to make sense of the $\delta$-correlated noise
by imposing an {\it ad-hoc} UV cutoff, one has no connection with
a realistic turbulence. On the contrary, $\alpha<-1$ is physical
as everything is related to the fluctuations on the largest
scales.
%
Flavor transformations in a medium with this spectrum of the
fluctuations, to the best of our knowledge, have not been treated
in the literature \footnote{Perhaps the closest is a perturbative
treatment of \cite{RashbaPRD} (essentially followed in
\cite{mubound}). It deals with neutrino spin-flavor precession in
the turbulent magnetic field not with density fluctuations that
are of interest here. Another notable reference is
\cite{Haxton:1990qb}, which deals with regular harmonic density
perturbations.}.

\section{Effects of fluctuations: a toy model}
We consider a 2-flavor system which goes through a ``noisy"
level-crossing, with a Kolmogorov spectrum. We take the
Hamiltonian
\begin{eqnarray}\label{eq:H2nuflavor}
  H_{2\nu}^{flavor} &=& \left(\begin{array}{cc}
  -\Delta\cos 2\theta + A(x)& \Delta\sin 2\theta \\
 \Delta\sin 2\theta &  \Delta\cos 2\theta - A(x)
\end{array}\right),
\end{eqnarray}
with $\Delta\equiv\Delta m^2/4 E_\nu$, $A(x)\equiv G_F
n_e(x)/\sqrt{2}$, and set $\Delta=50$, $\theta=0.1$,
$A(x)=x+A_{noise}(x)$. The noisy component of the profile is
modelled as
\begin{equation}\label{eq:numericalKolmogorov}
    A_{noise}(x)=F\sum_{k=1}^{600} k^{\beta} \cos[k x + \phi(k)],
\end{equation}
where $F$ is a normalization factor, $\beta=\alpha/2=-5/6$ and
$\phi(k)$ are random phases. The parameters are chosen such that
for $F=0$ the evolution is adiabatic (the adiabaticity parameter
$\gamma=\pi\Delta^2\sin^22\theta/|A'|\gg 1$), and the relevant
range $k \ge 2\Delta\sin2\theta=20$ (see later) is covered.

\begin{figure}[bth]
  \centering
  \includegraphics[width=0.48\textwidth]{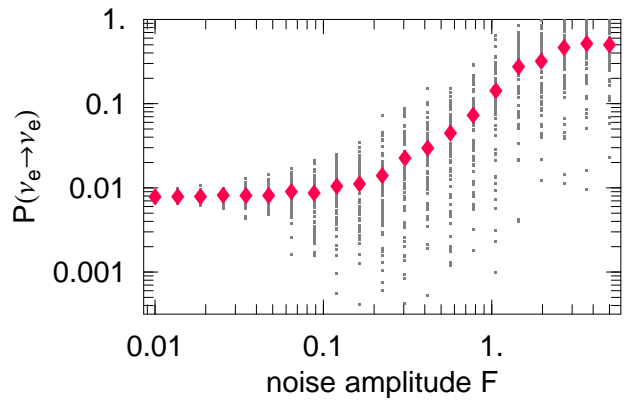}
\caption{The electron neutrino survival probability as a function
of the noise amplitude $F$ (see
Eq.~(\ref{eq:numericalKolmogorov})). For each $F$, the calculation
was repeated 66 times, with different random phases $\phi(k)$. The
points indicate the results of individual runs, while the diamonds
show the average values.} \label{fig:depolarization_log}
\end{figure}

Let us compute the survival probability $P$ of a neutrino of a
given flavor (for definiteness $\nu_e$) traveling from
$x_{start}=-100$ to $x_{end}=100$ for different values of the
noise amplitude $F$. Since the process is stochastic, for each $F$
we repeat the calculation with different random phases. The
results are shown in Fig.~\ref{fig:depolarization_log}. Three
different regimes are clearly seen: (i) For small noise
$(F\lesssim 10^{-2})$ its effects are negligible;
$P(\nu_e\rightarrow\nu_e)$ is dictated entirely by the smooth
component of the profile. (ii) As the noise is increased, a
stochastic scatter appears; from $F\sim 0.2$ to $1$ the average
survival probability -- shown by the diamonds -- grows \emph{as a
power law} with $F$. (iii) Finally, for $F\gtrsim 1-2$, the
average $P(\nu_e\rightarrow\nu_e)$ saturates to $1/2$; any value
of $P(\nu_e\rightarrow\nu_e)$ in the interval $[0,1]$ is equally
likely.

The saturation is easy to understand: in strongly fluctuating
backgrounds, the final state expectation value $\langle \nu_f
|\vec\sigma|\nu_f\rangle$ is equally likely to point in any
direction in the flavor space.
The averaged final state is described by a density matrix
$diag(1/2,1/2)$: it is completely depolarized, all information
about the initial state is lost. This result, well-known in the
case of the $\delta$-correlated noise, is expected in general.

\section{Method of analysis}
Fig.~\ref{fig:depolarization_log} suggests that to understand the
impact of the density noise on neutrinos
it is enough to derive the expression for
$P(\nu_e\rightarrow\nu_e)$ in the intermediate power-law regime.
If the answer turns out to be $\gtrsim1/2$, the noise completely
flavor-depolarizes the neutrino state and the true answer is
$P(\nu_e\rightarrow\nu_e)=1/2$. Given the significant
uncertainties in the explosion model, the problem may not warrant
anything more elaborate.


\section{Derivation of the effect of fluctuations}
To begin, consider a smooth density profile $n_0(x)$.
Rotate the Schr\"odinger equation, $i \phi'=H(x)\phi$, with $H(x)$
of Eq.~(\ref{eq:H2nuflavor}), to the basis $\psi(x)=V(x)\phi$
which diagonalizes the \emph{instantaneous} Hamiltonian. The full
Hamiltonian in the rotated basis has the form (see, {\it e.g.},
\cite{myresonance} for details)
\begin{eqnarray}\label{eq:H2numass}
  H_0^{mass} &=& \left(\begin{array}{cc}
  -\Delta_m& -i d\theta_m/dx \\
 i d\theta_m/dx &  \Delta_m
\end{array}\right),
\end{eqnarray}
where, introducing $A_0(x)\equiv G_F n_0(x)/\sqrt{2}$,
\begin{eqnarray}\label{eq:massquantities}
\Delta_m&=&[(\Delta\cos 2\theta-A_0)^2+\Delta^2\sin^22\theta]^{-1/2},\\
  \sin 2\theta_m &=& \Delta\sin 2\theta/\Delta_m.
\end{eqnarray}

Now let us add
a fluctuating density component $\delta n(x)$, so that $\delta
A(x)\equiv G_F \delta n(x)/\sqrt{2}$. We again rotate the
evolution equation to the instantaneous mass basis. Importantly,
however, we choose the basic defined not by the total matter term,
$A_0(x)+\delta A(x)$, \emph{but only} $A_0(x)$. We have
\begin{equation}\label{eq:schrodinger_deltaA}
    i \psi'=\left[H_0^{mass}(x) + \delta A\left(\begin{array}{cc}
 \cos 2\theta_m& \sin 2\theta_m \\
 \sin 2\theta_m &  -\cos 2\theta_m
\end{array}\right)\right]\psi.
\end{equation}

Transitions between the basis states are driven by the
off-diagonal terms $i d\theta_m/dx$ and $\delta A\sin 2\theta_m$.
Assuming the evolution without the fluctuations would be adiabatic
\footnote{It can be shown that in the opposite case the
fluctuations would not change the character of the
transition.}, we neglect the first of these terms and concentrate
on the second. Moreover, we assume $\Delta_m(x)\gg\delta A(x)\cos
2\theta_m$, i.e. approximate the diagonal splitting by
$2\Delta_m(x)$. Then, in the perturbative limit, the probability
of a transition is
\begin{eqnarray}\label{eq:P}
    P &\simeq& \left|\int_{x_i}^{x_f}dx \delta
    A(x)\sin2\theta_m(x) e^{i \int^x dx'2\Delta_m(x')}\right|^2.
\end{eqnarray}
Using Eq.~(\ref{eq:Cpowerlaw}), we get
\begin{eqnarray}\label{eq:Pgen_form}
    P &\simeq& \frac{G_F^2}{2}\int\frac{dk}{2\pi}C(k)\left|\int_{x_i}^{x_f}dx
    \sin2\theta_m(x)\right. \nonumber\\
     &\times&\left.\exp[i \int^x dx'(2\Delta_m(x)-k)]\right|^2.
\end{eqnarray}

Let us explore the implications of this result. First,
consider a turbulent density field superimposed on a constant
smooth density, $A_0(x)=const$. After the neutrino travels a
distance $L\gg\Delta_m^{-1}$, we have
\begin{eqnarray}
 \label{eq:Pconst}
P^{const}\simeq G_F^2 L \sin^22\theta_m C_0 (2\Delta_m)^\alpha/8.
\end{eqnarray}
The contributions come from a narrow interval of $k$, with the
width of $\sim L^{-1}$, centered on $k_0=2\Delta_m$.

Eq.~(\ref{eq:Pconst}) says that: (i) The depolarization is
efficient near the resonance, where $\sin2\theta_m$ is large and
$\Delta_m$ is small.
(ii) The effect is proportional to $L$ and may be significant even
off-resonance for a sufficiently extended turbulent region. We
will investigate both in turn.

Consider now a resonance crossing with a linear smooth component,
$A_0=A_0' x$. The integral over $x$ in Eq.~(\ref{eq:Pgen_form})
can be approximately evaluated using a stationary phase method.
For each $k>2\Delta\sin2\theta_{13}$ there are two stationary
points, which on physical grounds ({\it e.g.}, integration over
energy, variations in the profile) should be added incoherently.
We obtain
\begin{eqnarray}\label{eq:Plin_general}
P\simeq \frac{G_F} {\sqrt{2}n_0'}\int dk C(k)
G\left(\frac{k}{2\Delta \sin2\theta}\right),
\end{eqnarray}
where the spectral weight $G(p)$ is given by
\begin{eqnarray}
G(p)\simeq \Theta(p-1)p^{-1}(p^2-1)^{-1/2}.
\end{eqnarray}
The fluctuations that contribute have
wavelengths $\leq 4\pi E_{\nu}/(\Delta m^2 \sin2\theta)$, or, for the
$H$-resonance, $\leq
47$ km $\times(E_\nu/15$ MeV)$(0.3/\sin2\theta_{13})$.

For power-law $C(k)$ in Eq.~(\ref{eq:Cpowerlaw}) the integral in
Eq.~(\ref{eq:Plin_general}) can be evaluated analytically,
\begin{eqnarray}\label{eq:Ppower_law}
P^{power-law}\simeq\frac{G_F C_0}{\sqrt{2}|n_0'|}
(2\Delta\sin2\theta_{13})^{\alpha+1} \times N,
\end{eqnarray}
where the order one numerical coefficient $N$ is given by
$N=
\sqrt{\pi}\Gamma\left(1/2-\alpha/2\right)/[2\Gamma\left(1-\alpha/2\right)]$.

For the $\delta$-correlated noise ($\alpha=0$)
Eq.~(\ref{eq:Ppower_law}) gives
\begin{eqnarray}
P^{\delta-corr}\simeq\pi G_F C_0
\Delta\sin2\theta_{13}/\sqrt{2}|n_0'|.
\end{eqnarray}
This agrees with the known result, expanded in the same
perturbative limit. In the physically interesting case of the
Kolmogorov spectrum $(\alpha=-5/3)$, we get
$N=\sqrt{\pi}\Gamma(4/3)/2\Gamma(11/6)\approx 0.84$ and
\begin{eqnarray}\label{eq:PlinKolm}
P^{Kolm}\simeq 0.84G_F C_0 (2\Delta\sin2\theta_{13})^{-2/3}
/\sqrt{2}|n_0'|.
\end{eqnarray}
Eq.~(\ref{eq:PlinKolm}) reproduces the power-law rise seen in
Fig.~\ref{fig:depolarization_log}.

\section{Application to the supernova}
Let us apply Eq.~(\ref{eq:PlinKolm}) to the actual supernova case.
Four seconds into the explosion (\cite{Kifonidis:2005yj}, Fig.
4a), when the shock expands to $L\simeq5\times 10^4$ km, the
density in parts of the turbulent region drops to that of the $H$
resonance
\footnote{The two-flavor scheme is indeed applicable here: the
$H$-resonance occurs when \emph{two} levels, the $\nu_e$ (or
$\bar\nu_e$) and the heavy ``atmospheric" mass eigenstate
$\nu_\tau'\equiv\cos\theta_{23}\nu_\mu+\sin\theta_{23}\nu_{\tau}$
cross. In the interests of space, we refer the reader to
\cite{Smirnovandkids} for more complete definitions and a
discussion of the pre-shock signal.}.
Estimating $|n_0'|$ as $\sim n_0/L$ and using
$n_0=\sqrt{2}\Delta/G_F$ and $\delta n_L^2 \sim C_0
(2\pi/L)^{\alpha+1}/[2\pi|\alpha+1|]$,
we write the depolarization criterion ($P\gtrsim 1/2$, with $P$
from Eq.~(\ref{eq:PlinKolm})) as
\begin{equation}\label{depol}
{\delta n_L/ n_L}  > \tilde{f}~\theta _{13}^{-(\alpha+1)/2}(\Delta
L)^{-(\alpha+2)/2},
\end{equation}
where $\tilde{f}=\sqrt{\pi^\alpha/[N|\alpha+1|2^{\alpha+3}]}$ is
an order one factor.

For Kolmogorov's $\alpha=-5/3$, we get
\begin{equation}\label{depol1}
{\delta n_L/ n_L}  > 0.07~\theta _{13}^{1/3}.
\end{equation}
Fig.4a of \cite{Kifonidis:2005yj} shows that inside the turbulent
region $\delta n_L/n_L\sim$ a few. Thus, the depolarization
criterion is actually satisfied, and by a large margin.
Importantly, this result is robust to variations in the details of
the turbulent spectrum. As $\alpha$ is varied from $-2$ to $-1.5$,
the numerical factor in (\ref{depol1}) varies only from 0.04 to
0.25.


\section{Off-resonance depolarization}
By continuity, depolarization must be present even before the
density reaches its resonant value \footnote{The physics of this
off-resonance transition is essentially the parametric resonance
effect \cite{akhmedov}: the conversion is driven by wavenumbers in
the turbulent spectrum  that resonate with the inverse oscillation
length $2\Delta_m$.}. Let us estimate at what point the
depolarization effect becomes significant. One way is to model the
post-shock profile as a constant density region of extent $L\sim
\mbox{a few}\times 10^9$ cm.
Using Eq.~(\ref{eq:Pconst}) with $\sin
2\theta_m=\Delta\sin 2\theta/\Delta_m$, $G_F^2C_0/2\sim
A_0^2/k^{\alpha+1}=A_0^2 (L/2\pi)^{\alpha+1}$ and $\Delta_m\simeq
(A_0-\Delta)$, we get
\begin{equation}\label{eq:Pparametric_const}
    P^{const}\sim L^{\alpha+2}\Delta^2\sin^2 2\theta_{13} A_0^2
(A_0-\Delta)^{\alpha-2}/8 \pi^{\alpha+1}.
\end{equation}
Plugging in $E_\nu=1.5\times 10^{7}$ eV and $L\sim 2.5\times
10^{14}$ eV$^{-1}$ we find that $P=0.1$ is obtained when the
density is 4 times the resonant value for
$\sin^22\theta_{13}\sim0.1$; 2 times the resonant value for
$\sin^22\theta_{13}\sim0.01$; and 1.5 times the resonant value for
  $\sin^22\theta_{13}\sim0.001$.
To interpret these numbers, we consult, e.g., Fig. 2 of
\cite{RaffeltDighe} where contours of constant density are shown
as a function of time for a one-dimensional model of an exploding
supernova. We see that
for
$0.01\lesssim\sin^22\theta_{13}\lesssim0.1$ significant ($P\sim
0.1$) depolarization
occurs already at $t=3$ sec.

To check the robustness of this result to the details of the
profile, we can also model the smooth component $A_0(x)$ in the
post-shock region by a parabola, $A_0(x)=a+bx^2/2$. In the
far-off-resonance limit, $a-\Delta \gg \Delta\sin2\theta_{13}$, we
get from
Eq.~(\ref{eq:Pgen_form})
%
\begin{equation}\label{eq:Pparab_analyt}
    P^{parab} \simeq
 N_2\frac{\Delta^2\sin^22\theta_{13}}{\sqrt{b}}
 G_F^2C_0(a-\Delta\cos\theta_{13})^{\alpha-\frac{3}{2}},
\end{equation}
where
 $N_2=2^{\alpha-1/2}\sqrt{\pi}\Gamma(3/2-\alpha)/\Gamma(2-\alpha)$.
Eq.~(\ref{eq:Pparab_analyt}) yields an estimate that is quite
similar to the one obtained  using Eq.~(\ref{eq:Pparametric_const})
\footnote{Because the density close to the protoneutron star rises
rapidly, a better model may be just half of the parabola. This is
achieved by dividing the r.h.s. of Eq.~(\ref{eq:Pparab_analyt}) by
2.}.

\section{Experimental implications}
Let us briefly list the experimental implications of our findings.

(1) When the densities in the turbulence reach $\mbox{a few}\times
10^{3}$ g/cm$^3$, assuming $\sin^2 2\theta_{13}\gtrsim 10^{-3}$,
the $H$-resonance should become completely flavor-depolarized.
This manifests itself in a characteristic dip of the average
energy in the detector, or changing rate of the high-energy events. This is
qualitatively (but not quantitatively) similar to what has been
discussed in connection with the passage of the front shock
\cite{RaffeltDighe}. Just like in that case, the signature appears
either in neutrinos or antineutrinos, depending on the
sign of the neutrino mass hierarchy. An observation of the effect
would thus determine the sign of the mass hierarchy and place a
lower bound on $\theta_{13}$.

(2) The onset of the depolarization would be gradual, with the
off-resonance effects possibly present up to a second earlier.
This may be used to distinguish turbulence from the shock effects,
as the latter are more abrupt. The effect may be even more
pronounced for the $L$ resonance.

(3) The depolarization \emph{replaces} the signatures of the front
shock and other features in the density range covered by the
turbulence. Indeed, flavor-depolarized neutrino is described by a
density matrix $diag(1/2,1/2)$, which commutes with any
Hamiltonian and carries no information about density features
either before or after the turbulence. This extends to the Earth
effect when the $L$ resonance also becomes depolarized, providing
yet another possible way to distinguish turbulence from other
effects at future large detectors. The observation about the loss
of sensitivity to the front shock was also made in an interesting
recent paper \cite{Lisi2006}. Unfortunately, \cite{Lisi2006}
considers an {\it ad-hoc} $\delta$-correlated noise, following
\cite{Nicolaidis,Loreti1994,Loreti:1995ae,Balantekin1996,BurgessMichaud,BurgessProceedings}.

(4) The density profile varies between models and, moreover, even
within a given model \cite{Kifonidis:2005yj} varies significantly
depending on the direction to the observer. Consequently, there is no
unique prediction for when exactly the depolarization effects appear
first.

A detailed study of experimental signatures in various scenarios
(different hierarchies, ranges of $\theta_{13}$, explosion models)
will be published elsewhere.


\begin{acknowledgments}
We thank Evgeny Akhmedov, Sterling Colgate, Chris Fryer, George
Fuller, Wick Haxton, Thomas Janka, and Mark Wise for helpful
conversations. A.F. was supported by the US Department of Energy,
under contract number DE-AC52-06NA25396, A.G. by the David and
Lucile Packard Foundation.
\end{acknowledgments}


\begin{thebibliography}{99}

\bibitem{ColgateWhite1966}
  S.~Colgate and R.~White,
  Astrophys.\ J.\  {\bf 143}, 626 (1966).

\bibitem{Wilson1985}
J.~R.~Wilson, in {\it Numerical astrophysics}, edited by J.~M.~
Centrella, J.~M.~LeBlanc, and R.~L.~Bowers, Jones \& Bartlett,
Boston (1985).

\bibitem{Bethe1990}
  H.~A.~Bethe,
  Rev.\ Mod.\ Phys.\  {\bf 62}, 801 (1990).

\bibitem{HerantColgate1994}
  M.~Herant, W.~Benz, W.~R.~Hix, C.~L.~Fryer and S.~A.~Colgate,
  Astrophys.\ J.\  {\bf 435}, 339 (1994).

\bibitem{FryerWarren2003}
  C.~L.~Fryer and M.~S.~Warren,
  Astrophys.\ J.\  {\bf 601}, 391 (2004)
  [arXiv:astro-ph/0309539].

\bibitem{Kifonidis:2005yj}
  K.~Kifonidis, T.~Plewa, L.~Scheck, H.~T.~Janka and E.~Mueller,
  arXiv:astro-ph/0511369.

\bibitem{Scheck:2006rw}
  L.~Scheck, K.~Kifonidis, H.~T.~Janka and E.~Mueller,
  arXiv:astro-ph/0601302.

  \bibitem{JankaPRL}
  L.~Scheck, T.~Plewa, H.~T.~Janka, K.~Kifonidis and E.~Mueller,
  Phys.\ Rev.\ Lett.\  {\bf 92}, 011103 (2004).

\bibitem{Burrowsacoustic}
  A.~Burrows, E.~Livne, L.~Dessart, C.~Ott and J.~Murphy,
  arXiv:astro-ph/0510687.

\bibitem{WoosleyJanka}
  S.~Woosley and T.~Janka,
  Nature Physics {\bf 1}, 147 (2005).

\bibitem{HerantBenz1992} M.~Herant and W.~Benz,
Astrophys.\ J.\ {\bf 387}, 294 (1992).

\bibitem{SchiratoFuller}
  R.~C.~Schirato, G.~M.~Fuller,
  arXiv:astro-ph/0205390.

\bibitem{RaffeltDighe}
  R.~Tomas, M.~Kachelriess, G.~Raffelt, A.~Dighe, H.~T.~Janka and L.~Scheck,
  JCAP {\bf 0409}, 015 (2004).

\bibitem{Lisi2004}
  G.~L.~Fogli, E.~Lisi, A.~Mirizzi and D.~Montanino,
  JCAP {\bf 0504}, 002 (2005)
  [arXiv:hep-ph/0412046].

\bibitem{KnellerMcLaughlin}
  J.~P.~Kneller and G.~C.~McLaughlin,
  Phys.\ Rev.\ D {\bf 73}, 056003 (2006)
  [arXiv:hep-ph/0509356].

\bibitem{KamLANDspectrum2004}
  T.~Araki {\it et al.}  [KamLAND Collaboration],
  Phys.\ Rev.\ Lett.\  {\bf 94}, 081801 (2005)
  [arXiv:hep-ex/0406035].

   \bibitem{myMINOS}
  A.~Friedland and C.~Lunardini,
  arXiv:hep-ph/0606101.


\bibitem{myresonance}
  A.~Friedland,
  Phys.\ Rev.\ D {\bf 64}, 013008 (2001).



\bibitem{Nicolaidis}
  A.~Nicolaidis,
  Phys.\ Lett.\ B {\bf 262}, 303 (1991).

\bibitem{Loreti1994}
  F.~N.~Loreti and A.~B.~Balantekin,
  Phys.\ Rev.\ D {\bf 50}, 4762 (1994)
  [arXiv:nucl-th/9406003].

\bibitem{Balantekin1996}
  A.~B.~Balantekin, J.~M.~Fetter and F.~N.~Loreti,
  Phys.\ Rev.\ D {\bf 54}, 3941 (1996)
  [arXiv:astro-ph/9604061].

\bibitem{BurgessMichaud}
  C.~P.~Burgess and D.~Michaud,
  Annals Phys.\  {\bf 256}, 1 (1997)
  [arXiv:hep-ph/9606295].

\bibitem{BurgessProceedings}
  C.~P.~Burgess and D.~Michaud,
  arXiv:hep-ph/9611368.

\bibitem{Loreti:1995ae}
  F.~N.~Loreti, Y.~Z.~Qian, G.~M.~Fuller and A.~B.~Balantekin,
  Phys.\ Rev.\ D {\bf 52}, 6664 (1995).

\bibitem{Smirnovandkids}
  A.~S.~Dighe and A.~Y.~Smirnov,
  Phys.\ Rev.\ D {\bf 62}, 033007 (2000),
  hep-ph/9907423;
%
  C.~Lunardini and A.~Y.~Smirnov,
  JCAP {\bf 0306}, 009 (2003)
  hep-ph/0302033.

\bibitem{Haxton:1990qb}
  W.~Haxton and W.~Zhang,
    Phys.\ Rev.\ D {\bf 43}, 2484 (1991).

\bibitem{RashbaPRD}
  O.~G.~Miranda, T.~I.~Rashba, A.~I.~Rez and J.~W.~F.~Valle,
  Phys.\ Rev.\ D {\bf 70}, 113002 (2004)
  [arXiv:hep-ph/0406066].

  \bibitem{mubound}
   A.~Friedland,
  arXiv:hep-ph/0505165.

  \bibitem{akhmedov}
  E.~Kh.~Akhmedov, Yad.\ Fiz.\ {\bf 47} 475 (1988) [Sov.\ J.\
  Nucl.\ Phys. {\bf 47} 301 (1988)].

\bibitem{Lisi2006}
  G.~L.~Fogli, E.~Lisi, A.~Mirizzi and D.~Montanino,
  arXiv:hep-ph/0603033.

\end{thebibliography}
\end{document}